\documentclass[10pt,journal,compsoc]{IEEEtran}
\usepackage{blindtext}
\usepackage{hyperref}

\ifCLASSOPTIONcompsoc
   \usepackage[nocompress]{cite}
\else
    \usepackage{cite}
\fi

\ifCLASSINFOpdf
 
\else
 
\fi

\usepackage{graphicx}
\usepackage{xcolor}

\begin{document}

\title{Privacy-Preserving Cloud Computing: Ecosystem, Life Cycle, Layered Architecture and Future Roadmap}
\author{\IEEEauthorblockN{ \textbf{Saeed Ahmadi\IEEEauthorrefmark{1}}  }\\
	\IEEEauthorblockA{\textit{\IEEEauthorrefmark{1}School of Computer Science,} 
		\textit{University of Guelph, Ontario, Canada  }\\
		sahmad18@uoguelph.ca}
	\\
	\and
	\IEEEauthorblockN{ \textbf{Maliheh Salehfar\IEEEauthorrefmark{2}}}\\
	\IEEEauthorblockA{\textit{\IEEEauthorrefmark{2}School of Management and Accounting} \\
		\textit{Allameh Tabataba'i University, Tehran, Iran }\\
		malihehsalehfar@gmail.com}}




\IEEEtitleabstractindextext{%
\begin{abstract}
\textcolor{black}{Privacy-Preserving Cloud Computing is an emerging technology with many applications in various fields. Cloud computing is important because it allows for scalability, adaptability, and improved security. Likewise, privacy in cloud computing is important because it ensures that the integrity of data stored on the cloud maintains intact.  This survey paper on privacy-preserving cloud computing can help pave the way for future research in related areas. This paper helps to identify existing trends by establishing a layered architecture along with a life cycle and an ecosystem for privacy-preserving cloud systems in addition to identifying the existing trends in research on this area. }
\end{abstract}

\begin{IEEEkeywords}
Cloud Computing, Privacy, Secure Cloud, Privacy-Preserving Cloud, Trend Analysis, Future Roadmap.
\end{IEEEkeywords}}

\maketitle

\IEEEdisplaynontitleabstractindextext

\IEEEpeerreviewmaketitle

\section{Introduction}\label{sec:introduction}

In recent years, cloud computing has been of great interest to the research community \cite{TYV-Jour001}\cite{TYV-Jour002}. It is used in a variety of applications ranging from healthcare \cite{Intro-Jour001}\cite{Intro-Jour002} to geosicences \cite{Intro-Jour004} and autonomous vehicles \cite{Intro-Jour003,a1}. Cloud computing plays a significant role in modern technology as it is actively interacting with other technological fields such as edge \cite{Intro-Jour007} and fog computing \cite{Intro-Jour008}, \textcolor{black}{Internet of Things (IoT)} \cite {Intro-Jour005}, sensor technology \cite{Intro-Jour006} and big data \cite{Intro-Jour009}\cite{Intro-Jour010}. \textcolor{black}{"Edge Computing has been defined as a distributed computing topology in which the storage and computing has been perfume close to the source of data.Fog Computing has been defined as decentralized computing infrastructure that perfume between the data source and the cloud.} Enabling technologies that support cloud computing vary from real-time computing \cite{Intro-Jour013} and virtualization \cite{Intro-Jour017} to artificial intelligence \cite{Intro-Jour014}\cite{Intro-Jour015}\cite{Intro-Jour016}. Several design objectives including performance \cite{Intro-Jour024}, reliability \cite{Intro-Jour025} and fault tolerance \cite{Intro-Jour023} are considered in the design of cloud computing systems. Especially, concerned about different aspects of security including attack resilience \cite{Intro-Jour028,abc}\cite{Intro-Jour027} and confidentiality \cite{Intro-Jour029}\cite{Intro-Jour030} have been of concern to cloud computing researchers. However, privacy seems to be the most critical security aspect in cloud computing \cite{TBT-Jour001}\cite{TBT-Jour002}\cite{a2}.

\textcolor{black}{ In this paper, we first establish an ecosystem for privacy-preserving cloud including applications, vehicular technology, healthcare, social networking, Internet of Things (IoT), smart home, big data, and data centers. Then we developed a life cycle and reviewed it in these stages: design, verification, implementation, and deployment. In the next step, we build a layered architecture containing Privacy-Preserving Infrastructure, Privacy-Preserving Services, and Privacy-Preserving Computing. Lastly, we establish a future roadmap in Quantum-Inspired and Bio-Inspired AI. We present our paper with the hope that we can motivate others to study these topics further as these new technologies develop.}
\textcolor{black}{It should be noted that although there may be existing survey papers related to the topic at hand, they have shortcomings which motivate our work in this survey. It is our intention to cover the topic of Privacy-Preserving Cloud Computing as comprehensively as possible. }

The rest of this paper is organized as follows. Section 2 summarizes existing surveys. Section 3 outlines the applications and technologies that comprise the ecosystem for cloud computing. Section 4 outlines the full 4-stage life cycle of a cloud system . Section 5 describes the layers which make up the architecture of a privacy-preserving cloud system. Finally, Section 6 outlines a future roadmap for cloud computing and how \textcolor{black}{Artificial Intelligence (AI) }can be used to create new technologies.

\begin{figure}[h]
	    \centering
	    \includegraphics[width=0.9\linewidth,keepaspectratio]{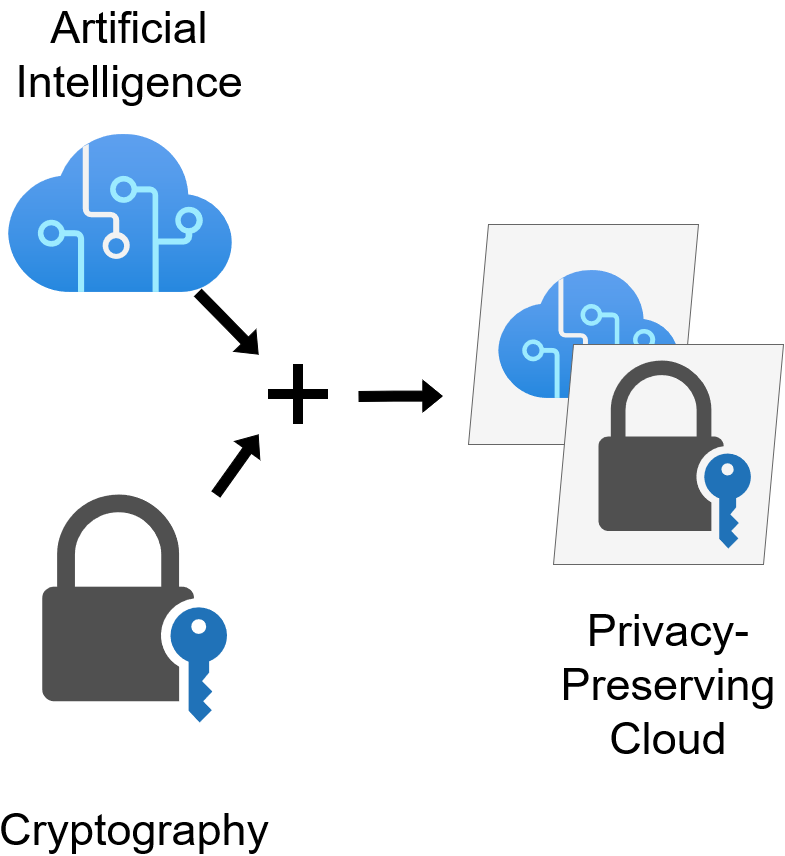}
	    \caption{Privacy Incorporated in the Design of Cloud Computing Environments}
	    \label{MyFig1}
\end{figure}

\textcolor{black}{Privacy is important in cloud computing because it ensures that data shared in the cloud can only be accessed by those who have been given permission. Without proper data privacy, the integrity of data stored in the cloud would not be able to be guaranteed. In Figure \ref{MyFig1}, Artificial Intelligence and Cryptography combine together, resulting in a Privacy-Preserving Cloud system where privacy is built into the foundation using AI and Cryptography concepts.}

\textcolor{black}{ In Figure \ref{MyFig1}, the blue cloud represents Artificial Intelligence and the Lock and Key represent Cryptography. The arrows leading from these symbols point towards a plus symbol, signifying their union. The result of this combination is Privacy-Preserving Cloud, represented by the joining of the two symbols.}

\section{Existing surveys}\label{ExSurv}

There are many surveys on the security of cloud computing. However, some of them are too outdated for such a dynamic research area. Others do not focus on the privacy of cloud computing \cite{a3}. Some existing surveys study cloud computing only in specific applications, and some of them fail to develop a future roadmap. These shortcomings motivate our work in this paper.

\subsubsection{Surveys on Challenges, Issues, Concerns and Requirement of Cloud Computing}

Storing data deals with many challenges in security and privacy, and Cloud Service Providers should take appropriate steps to overcome these challenges \cite{a4}. In this section, previous studies and outputs have been reviewed. Although Mobile Cloud Computing (MCC) provides a way to meet our needs for abundant resources in transportable devices, there is a lack of security and confidentiality. To that end, in \cite{Surv-Conf036}  security issues and challenges in mobile cloud computing have been analyzed. In \cite{Surv-Conf037} Data Integrity (DI), Data Confidentiality (DC), and availability that are very important in cloud storage have been studied. Furthermore, one of the famous and popular technology for organizations and personal users that Cloud Computing has provided is Storage  as  a  Service(StaaS). The security and privacy challenges according to growing concern in STaaS have been presented in \cite{Surv-Conf038}. Decreasing security risks in Cloud Computing is the fundamental action that has been analyzed in \cite{Surv-Conf039}, \cite{Surv-Conf040} that authors proposed solutions to overcome. These concerns were further studied in \cite{Surv-Conf041} and \cite{Surv-Conf042} since that is not obvious where the data is stored and controlled. In \cite{Surv-Conf043}, the Pixel key pattern and Image Steganography techniques have been proposed to solve insuperable data security problems in Cloud Computing but in \cite{Surv-Conf044} potential issues for gathering access to users data and harming their privacy have been studied in order to attract urgent attention. Three threats are data breaches,   hijacking of accounts, and multitenancy that brought about security issues that have been analyzed in \cite{Surv-Conf045}, and predicted that if these issues have been solved, cloud computing will be more widespread. Although The requirements and solutions of cloud computing security were reviewed comprehensively in \cite{Surv-Conf046}, its publication year is old, and its solution may not be helpful for present technology according to fast development. In \cite{Surv-Conf047} and \cite{Surv-Conf048}, similar algorithms have been proposed to overcome the obstacles in Cloud Computing under three fundamental categories, cryptographic, data storage, and data semantics. Moreover, Information that users should be aware of to estimate risks related to preserving their data was presented in \cite{Surv-Conf049}, but the privacy of users was not under observed.

\subsection{Surveys on the Evolution and Future Foresight}
Cloud Computing and Blockchain technology development have been examined in \cite{a5,a6}, and how these technologies can integrate and improve each other drawbacks were under observation. 

\subsection{Surveys on the Applications of Cloud}

\subsubsection{Applications in Healthcare and Medical Systems}
In \cite{Surv-Conf059}, the authors argued the security peril of cloud service platforms in the field of medicine and also provided related solutions. Protecting privacy and security of users and data were presented in \cite{Surv-Conf059} and explained a method that can refuse illegal users permission.

\subsection{Surveys on Secure Cloud}
The existing surveys related to infrastructure and the concept of security in Cloud Computing have been reviewed in this section. Likewise, in the first subsection of this section, we investigate the general security concept in Cloud Computing, and in the subsequent subsections, privacy and security of data, storage, communication \cite{c6} have been studied. Moreover, cryptographic and steganography methods, \textcolor{black}{and} Secure Data Sharing risks have been reviewed in current surveys subsequently. Following \textcolor{black}{sections}, authentication and security auditing, as well as data provenance and challenges and risks, threats and attacks, have been investigated.

\subsubsection{The General Concept of Security}
In \cite{Surv-Conf060} authors came up with a method for solving the problem related to enterprise users, cloud operators, and regulators by analyzing current security and privacy challenges, in addition in  \cite{Surv-Conf061}, authors have analyzed the solution of different categories of risks in cloud systems such as infrastructure threats, host threats, and service providers threats similar to \cite{Surv-Conf062} that authors have introduced a solution to overcome security issues related to confidentiality, integrity, availability, authentication. In comparison to \cite{Surv-Conf063}, where the authors have viewed the privacy and security of data from the perspective of appropriate training because the consultant is a crucial part of service providers that are dealing with the implementation. The authors in \cite{Surv-Earl-Jour003} have observed the effect of extended cloud such as \textcolor{black}{Mobile Edge Computing (MEC)} and fog on network schematic of cloud computing; also, they analyzed models and architectures of technologies due to the security and flexibility that is required while in \cite{Surv-Earl-Jour004}, the authors have analyzed the algorithms that come from biological In order to deal with security issues of cloud computing. In addition, the authors in \cite{Surv-Conf065} have come up with an innovative method which is Cloud Bursting Brokerage and Aggregation (CBBA), and analyzed the secure sharing mechanism to reduce cloud-bursting with denial-of-service (DoS) attacks or distributed denial-of-service (DDoS) attacks that have deleterious effects on cloud computing.

As opposed to \cite{Surv-Conf066} that Secure Smart Grids in cloud-based software platforms was observed, the authors in \cite{Surv-Conf067} have explained the model, control, and management of the rechargeable batteries such as Li-ion batteries used in IoT. Furthermore, the vulnerabilities and threats in the live migration of Virtual Machine have been presented in \cite{Surv-Conf068} and claimed that there is an inadequate solution to provide security. Moreover, \textcolor{black}{Sun and Aida} have indicated how they meet the security of user applications with a method for running applications that need huge resources on both local computing resources and Infrastructure-as-a-Service Cloud (IaaS) by making use of machine learning in \cite{Surv-Conf069} similar to \cite{Surv-Conf070} where the authors have demonstrated how they came over several basic issues and improved CPU frequency in Virtual Machine with Workflow Scheduling. In addition, the security of live migration in virtualization has been analyzed in \cite{Surv-Conf071}, and authors have introduced three techniques to cover confidentiality and integrity of data, authentication, and authorization of operations.

\subsubsection{Privacy-Preserving Cloud}
A framework of privacy protection in cloud computing has been demonstrated in \cite{Surv-Earl-Jour006}, and the authors have analyzed the concept of encryption and security and draw future directions, while the authors in \cite{VBAGH-Jour001} reviewed a number of articles related to the privacy and security of electronic health records in cloud architecture and the cryptographic and noncryptographic of electronic health records protection. In addition, the vulnerabilities and threats in Cloud Computing have been analyzed in terms of confidentiality, integrity, availability, accountability, and privacy-preservability in \cite{Surv-Earl-Jour008} and they have proposed strategies for protection. Also, for dealing with challenges and improving the level of privacy in Cloud Computing, Privacy Enhancing Technologies was studied by authors in \cite{Surv-Earl-Jour009} in order to identify a method. The authors have demonstrated a rundown of technologies and patterns in order to protect and encrypt data in \cite{Surv-Earl-Jour010} and emphasized issues and tackles that can be studied in the future in Vehicular Cloud Computing in \cite{Surv-Earl-Jour011} by reviewing and examing the structure and concept of cloud storage and adventures and risks of security and privacy of cloud storage. Although, the authors in \cite{Surv-Earl-Jour012} argued the most advanced method of privacy-preserving that is utilized in e-Health clouds as well as explained categorization of cryptographic and noncryptographic approaches, in \cite{Surv-Earl-Jour014} the security and privacy of health data stored in the cloud have been reviewed in order to indicate the advantages and disadvantages of eHealth's security and privacy algorithms. Several implementation techniques that can help developers and a number of privacy issues that should be attended to in the could systems environment have been presented by authors in \cite{Surv-Earl-Jour015}. As opposed to \cite{Surv-Conf075} where the number of analyses that carried out in cloud computing has been examined in order to make know threats related to cloud computing architecture, authors in \cite{Surv-Conf076} argued security and privacy issues in order to provide appropriate solutions. Moreover, an encryption data method which is the Bloom filter technique has been presented in \cite{Surv-Conf077}.

\subsubsection{Secure Data}

Data Security is vital and important in cloud computing \cite{c7}. The encryption algorithms have been analyzed in terms of providing the security of data in storing and transferring in \cite{Surv-Conf001} similar to \cite{Surv-Conf002} in order to overcome the challenges in the security of information. In \cite{Surv-Conf003}, the perceptions of audit and data security that is the fundamental obstacle that should pay attention to having been analyzed, while in \cite{Surv-Conf004} Multi-Cloud Database (MCDB) has been reviewed, and they explained its drawback, which is the multi-stage cloud that should be considered in order to make it simple in the future. Furthermore, cloud services such as \textcolor{black}{Software as a Service (SaaS), Platform as a Service (PaaS), Infrastructure as a Service (IaaS)} have been analyzed in terms of cloud computing security issues in \cite{Surv-Conf005}, and also the authors provided some solutions for security and overcome its drawbacks. In \cite{Surv-Conf006}, the Mobile Cloud Computing data security frameworks have been under observation to recognize obstacles, issues, and security measures for improvement in future implementation. Moreover, encryption and decryption algorithms have been presented in \cite{Surv-Conf007} and authors have proposed new algorithms that improve the security and efficiency of data in cloud computing.

\paragraph{Data Security Standards}\mbox{}
The existing method of integrity, authentication, and confidentiality of data in cloud computing has been presented in \cite{Surv-Conf008} and \cite{Surv-Conf009}; also the authors analyzed security techniques, issues, and challenges to improve data security measures in cloud computing.

\subsubsection{Secure Storage}

There is growing concern about storing data in the cloud \cite{a7}. In \cite{Surv-Conf012} the authors have demonstrated how the secret key, used in encryption and decryption, has affected the efficiency and security of bandwidth and storage space. The security of data storage related to diabetes disease in cloud computing was under observation in \cite{Surv-Conf010}, and authors brought up some issues that need to be considered for the framework that collects information similar to  \cite{Surv-Conf011} where several cloud data storage infrastructure and security algorithms were presented and offered the accountability process, and protection measures that need to be developed in further study. In addition, the effects of data security in cloud storage on integrity, confidentiality, and availability were analyzed in order to recognize security requirements \textcolor{black}{that} have been analyzed in \cite{Surv-Conf014}. Also, encryption and decryption algorithms used in cloud storage were analyzed and compared with related techniques to improve their performances in \cite{Surv-Conf015}. Authors have claimed sharing operation and searching operation architect needs to develop and design securely in the future. In contrast, in \cite{Surv-Conf016} challenges and limitations were observed in terms of data security in these systems and authors have argued how security and usage of information provided in various search systems in cloud storage similar to \cite{Surv-Conf017} that authors have analyzed security algorithms and techniques used in the client-side and vendor-side. Moreover, The advantages of using blockchain in cloud computing were presented in \cite{Surv-Conf018} as well as authors analyzed how these methods improve the security of data in cloud computing due to its decentralized nature. Likewise, The authors in \cite{Surv-Conf019} have analyzed security data and provided solutions to improve third parties infrastructure in terms of safety, privacy, and availability measures.

\subsubsection{Secure Communication}

For reviewing secure communication subjects in current surveys, \textcolor{black}{Soni and Kumar} have analyzed methods and algorithms related to cloud computing security to find the gap between future development and previous research, and they gave prominence to future suggestions.

\subsubsection{Cryptographic Security}
Cryptographic Security is one of the fundamental methods for providing security and privacy of data in Cloud Computing \cite{a8}. Proxy re-encryption, which can encrypt data in cloud computing with user private key, has been under observation in \cite{Surv-Earl-Jour001} and authors, by comparing it with current algorithms and measures have introduced better security and efficient data solutions similar to \cite{Surv-Earl-Jour002} where the security schema and design of public-key encryption with equality test \textcolor{black}{ (PKE-SF)} was discussed and authors have examined its feature efficiency and proposed some helpful directions. In addition, various cryptographic techniques in the public cloud have been reviewed in \cite{NHYU-Jour001}, and demonstrated how these techniques could preserve data from malicious activity in cloud computing and highlighted their issues similar to \cite{Surv-Conf021} that several security issues were presented, and authors proposed various solutions for the purpose of getting to know about them for increasing security in cloud conditions. Furthermore, in \cite{Surv-Conf022} and \cite{Surv-Conf023} the authors compared several cryptographic methods and demonstrated differences in cost and performance, and genetic algorithms and visual cryptography have been proposed for concealing high-priority data similar to \cite{Surv-Conf024} that authors have proposed cryptographic techniques that improve the security of cloud storage.

\subsubsection{Steganography}
In \cite{Surv-Conf025} authors have described its related techniques: the Least Significant Bit (LSB) and the Discrete Cosine Transform (DCT).

\subsubsection{Secure Data Sharing}
The security of data sharing is one of the most critical steps in Cloud Computing \cite{c5}. Authors in \cite{Surv-Conf026} have introduced several methods that provide data sharing security in the cloud environment but mainly analyzed Ciphertext Policy Attribute-Based Encryption(CP-ABE) \textcolor{black}{that has been}used in integrity policies.

\subsubsection{Risks, Threats and Attacks}
In \cite{Surv-Non-Earl-Jour001}, the authors reviewed several papers and identified several security issues that affected the cloud computing environment. They conclude that the blockchain mitigates security concerns for service providers and users similar to \textcolor{black}{\cite{Surv-Conf027} }that authors have proposed Advanced Security Measures and Practices to improve the structure of current technology to deal with security risks and challenges. The structure of the blockchain method has been analyzed in \cite{Surv-Conf028}, and the authors proposed a secure method of blockchain that can overcome security issues and risks as opposed to \cite{Surv-Conf029} where the authors have explained the vulnerabilities, attacks, and threats related to cloud services and infrastructures. In addition, the benefits and drawbacks of current security methods have been analyzed in \cite{Surv-Conf030}, and the authors argued potential capabilities in the security of cloud service in order to overcome disadvantages in comparison to \cite{Surv-Conf031} that authors did a thorough investigation in the security of cloud infrastructure and offered a method that can improve issues and risks in that time. However, due to the year when the paper was conducted, this method would not be effective. In addition, In \cite{Surv-Conf032}, authors conducted a detailed analysis on papers between 2019 and 2020 to help companies be aware of cloud computing vulnerabilities and prioritize due to the widespread use of cloud services as opposed to in \cite{Surv-Conf033} that risk management methods that are different from those in IT enterprise have been analyzed in cloud systems and authors have demonstrated its efficiency in cloud security challenges by observing technically. The accuracy of \textcolor{black}{ Intrusion Detection Systems (IDS) } have been analyzed in cloud base malicious, and attacks in \cite{Surv-Conf034} and authors have reported IDS which utilizes multiple method detection to improve its efficiency in the cloud while in \cite{Surv-Conf035} the current security framework has been analyzed to accentuate the disadvantages and drawbacks and authors have implemented new \textcolor{black}{ Intrusion Detection Systems(IDS) } and Prevention System (IPS) that improve security in virtual networks.

\subsubsection{Authentication}
In \cite{Surv-Conf050}, the authors have made use of Kerberos as a third-party auditor for authentication, RSA algorithm \textcolor{black}{(Rivest–Shamir–Adleman)} for security, and MD5 algorithm to provide integrity and also how the cloud systems provide security when information data is being transferred. In addition, in \cite{Surv-Conf051}; the duplication process has been presented in detail, and new techniques in a cloud architecture that encrypts duplicate-check tokens of files with private keys have been presented.

\subsubsection{Data Provenance}
The analysis of current provenance management schemes in the cloud system and their security solution was presented in \cite{Surv-Conf052}.

\subsubsection{Security Auditing}
The advantages and drawbacks of several Third Party Auditing (TPA) have been analyzed in \cite{Surv-Conf053}, and the authors have concluded that nearly all TPA methods need to be developed in terms of the security of public auditing.

\subsection{Surveys on Cloud-Based Big Data}
Big Data refers to an enormous amount of data that has been produced these days by different sources, Such as IoT devices. Since storing a vast amount of data in Cloud Computing is becoming a challenge for service providers. Different security methods of eHealth data in the cloud have been presented in \cite{Surv-Conf054}, also in order to improve security, and the authors came up with a decoy technique that is storing the patient data initially in the cloud and another in the fog layer. In contrast, in \cite{Surv-Conf055}, prior research in the security of storage management has been analyzed, and authors have proposed a new technique.

\subsection{IoT-Based Cloud and Cloud-Based IoT}
Internet of Things \textcolor{black}{(IoT)} devices is widespread these days since there is no doubt in the benefits of those devices, but when it comes to private data, the security mechanism has played a vital role \cite{a9}. This section studied IoT and cloud computing, two concepts that have become so interwoven and almost inseparable. Several infrastructures of IoT were analyzed effectively from the perspective of security in \cite{Surv-Conf056}, and authors have observed them by conducting the man-in-the-middle attack between end devices sensor and service providers while in \cite{Surv-Conf057} just the security of data in the Internet of Things (IoT) integrated with Cloud base internet has been analyzed, and risks and threats in the Internet of Things have been highlighted. Moreover, the security risks related to storing and collecting data from IoT devices that need urgent action have been presented in \cite{Surv-Conf058}. In addition, the authors came up with solutions that may be helpful as opposed to \cite{Surv-Earl-Jour002-1} that threats, risks, vulnerabilities of \textcolor{black}{ Supervisory Control And Data Acquisition} (SCADA) systems have been analyzed, and some crucial issues reported. WebSCADA is an application utilized for innovative medical technologies.


\subsection{Surveys on AI-Assisted Secure Cloud}
The authors reviewed 60 articles in order to demonstrate challenges and accuracy estimation of using machine learning and the type of machine learning in cloud security \cite{a10}. The efficiency of facial recognition techniques in a cloud environment from data security has been analyzed in \cite{TYBH-Conf001}. Cloud data behavior detection and recognition are the major components that will contribute to developing a secure data surveillance system implemented in a cloud environment that will protect stored data from network intrusions and ensure data security.

In this location, you need a table. Each row of the table is allocated to one of the above surveys. The first coulmn in each row cites the related survey paper. The second row contains the publication year of the survey. The third row indicates whether or not the survey focuses on the privacy of cloud computing (Yes/No). The fourth column Indicates whether or not the survey presents a future roadmap (Yes/No). The fifth column indicates whether or not the survey studies cloud computing in a specific application (Yes/No).

The summary of previous studies related to Privacy-Preserving in Cloud Computing has been presented in  \textcolor{black}{Table \ref{MyTable1} and Table \ref{MyTable2}}. \textcolor{black}{Table \ref{MyTable1}} includes from \cite{Surv-Conf036} to \cite{Surv-Conf015} and \textcolor{black}{Table \ref{MyTable2}} from \cite{Surv-Conf016} to \cite{TYBH-Conf001}. The second column refers to the published year, and the third column shows whether the paper focuses on the privacy of cloud computing or not. The next column indicates whether the paper creates a future road map or not, and the last column specifies if they have been studied in a specific application. In both tables, the papers have been reviewed as 2010. In the \textcolor{black}{Table \ref{MyTable1}}, just 13 papers focus on privacy in cloud computing, while in \textcolor{black}{Table \ref{MyTable2}}, no paper focuses on privacy in cloud computing. Some papers have drawn a future road map and analyzed cloud computing in a specific application.


\begin{table}
\caption{Summary of previous studies in Privacy-Preserving Cloud Computing - Part 1}
\label{MyTable1}
  \begin{center}
  \begin{tabular}{| p{0.12\linewidth} | p{0.055\linewidth} | 
p{0.19\linewidth} | p{0.11\linewidth} | p{0.14\linewidth} |}
   \hline
  Published & Year &NN-focuses on the privacy of cloud computing & a future roadmap
&  cloud  computing  in  a  specific  application  \\ [0.7ex]
  \hline
   \cite{Surv-Conf036} & 2018 & no & no & no\\
   \hline
   \cite{Surv-Conf037} & 2019 & no & no & no \\
   \hline
   \cite{Surv-Conf038} & 2019 & no & no & yes \\
   \hline
   \cite{Surv-Conf039} & 2020 & yes & no & no  \\
   \hline
   \cite{Surv-Conf040} & 2015 & no & yes & no \\
   \hline
   \cite{Surv-Conf041} & 2015 & no & yes & no \\
   \hline
   \cite{Surv-Conf042} & 2015 & yes & yes & no \\
   \hline
   \cite{Surv-Conf043} & 2015 & yes & yes & yes \\
   \hline
   \cite{Surv-Conf044} & 2014 & yes & no & no \\
   \hline
   \cite{Surv-Conf045} & 2016 & no & no & no \\
   \hline
   \cite{Surv-Conf046} & 2012 & no & yes & no \\
   \hline
   \cite{Surv-Conf047} & 2017 & no & yes & yes \\
   \hline
   \cite{Surv-Conf048} & 2020 & no & yes & no \\
   \hline
   \cite{Surv-Conf049} & 2015 & no & no & no \\
   \hline
   \cite{Surv-Conf064} & 2018 & no & no & yes \\
   \hline
   \cite{Surv-Conf059} & 2013 & no & no & yes \\
   \hline
   \cite{Surv-Conf060} & 2013 & no & no & no \\
    \hline
   \cite{Surv-Conf061} & 2017 & no & no & no \\
   \hline
   \cite{Surv-Conf062} & 2016 & no & no & no \\
    \hline
   \cite{Surv-Conf063} & 2017 & no & no & yes \\
   \hline
   \cite{Surv-Earl-Jour003} & 2020 & no & yes & yes \\
   \hline
   \cite{Surv-Earl-Jour004} & 2011 & no & no & yes \\
   \hline
   \cite{Surv-Conf065} & 2014 & no & no & no\\
   \hline
   \cite{Surv-Conf066} & 2017 & no & no & yes\\
   \hline
   \cite{Surv-Conf067} & 2013 & no & no & no\\
   \hline
   \cite{Surv-Conf068} & 2013 & no & no & yes\\
   \hline
   \cite{Surv-Conf069} & 2010 & no & no & yes\\
   \hline
   \cite{Surv-Conf070} & 2018 & no & no & yes\\
   \hline
   \cite{Surv-Conf071} & 2014 & no & no & yes\\
   \hline
   \cite{Surv-Earl-Jour006} & 2019 & yes & yes & no\\
   \hline
   \cite{VBAGH-Jour001} & 2019 & yes & no & yes \\
    \hline
   \cite{Surv-Earl-Jour008} & 2012 & yes & no & no \\
    \hline
   \cite{Surv-Earl-Jour009} & 2021 & yes & yes & no \\
    \hline
   \cite{Surv-Earl-Jour010} & 2020 & yes & no & no \\
    \hline
   \cite{Surv-Earl-Jour011} & 2020 & yes & yes & yes \\
    \hline
   \cite{Surv-Earl-Jour012} & 2014 & yes & yes & yes \\
    \hline
   \cite{Surv-Earl-Jour014} & 2021 & yes & yes & yes \\
    \hline
  \cite{Surv-Earl-Jour015} & 2014 & yes & no & no \\
    \hline
   \cite{Surv-Conf075} & 2017 & no & no & no \\
    \hline
   \cite{Surv-Conf076} & 2016 & no & no & no \\
    \hline
   \cite{Surv-Conf077} & 2018 & no & no & no \\
   \hline
   \cite{Surv-Conf001} & 2019 & no & no & no \\
    \hline
   \cite{Surv-Conf002} & 2021 & no & no & no \\
    \hline
   \cite{Surv-Conf003} & 2018 & no & no & no \\
    \hline
   \cite{Surv-Conf004} & 2020 & no & yes & yes \\
   \hline
   \cite{Surv-Conf005} & 2021 & no & no & no\\
   \hline
   \cite{Surv-Conf006} & 2019 & no & yes &yes\\
   \hline
   \cite{Surv-Conf007} & 2020 & no & no & no\\
   \hline
   \cite{Surv-Conf008} & 2020 & no & no & no\\
   \hline
   \cite{Surv-Conf009} & 2018 & no & no & no\\
   \hline
   \cite{Surv-Conf010} & 2017 & no & no & yes\\
   \hline
   \cite{Surv-Conf011} & 2020 & no & yes & no\\
   \hline
   \cite{Surv-Conf012} & 2015 & no & no & yes\\
  \hline
   \cite{Surv-Conf014} & 2019 & no & no & no\\
  \hline
   \cite{Surv-Conf015} & 2021 & no & yes & no\\
     \hline
  \hline
  \end{tabular}
  \end{center}
\end{table}

\begin{table}
\caption{Summary of previous studies in Privacy-Preserving Cloud Computing - Part 2}
\label{MyTable2}
  \begin{center}
  \begin{tabular}{| p{0.12\linewidth} | p{0.055\linewidth} | 
p{0.19\linewidth} | p{0.11\linewidth} | p{0.14\linewidth} |}
   \hline
   Published & Year & NN-focuses on the privacy of cloud computing & a future roadmap
&  cloud  computing  in  a  specific  application  \\ [0.7ex]
  \hline
    \cite{Surv-Conf016} & 2021 & no & no & no\\
   \hline
   \cite{Surv-Conf017} & 2020 & no & no & no \\
   \hline
   \cite{Surv-Conf018}& 2020 & no & no & yes \\
   \hline
   \cite{Surv-Conf019}& 2019 & no & no & no  \\
   \hline
   \cite{Surv-Conf020} & 2017 & no & yes & no \\
   \hline
   \cite{Surv-Earl-Jour001} & 2016 & no & yes & yes \\
   \hline
   \cite{Surv-Earl-Jour002} & 2021 & no & no & yes \\
   \hline
   \cite{NHYU-Jour001} & 2016 & no & no & no \\
   \hline
   \cite{Surv-Conf021} & 2021 & no & no & no \\
   \hline
   \cite{Surv-Conf022} & 2014 & no & no & no \\
   \hline
   \cite{Surv-Conf023} & 2017 &  no & no & no \\
   \hline
   \cite{Surv-Conf024} & 2014 & no & no & yes \\
   \hline
   \cite{Surv-Conf025} & 2019 & no & no & yes \\
   \hline
   \cite{Surv-Conf026} & 2021 & no & no & yes \\
   \hline
   \cite{Surv-Non-Earl-Jour001} & 2021 & no & yes & no \\
   \hline
   \cite{Surv-Conf027} & 2020 & no & no & no \\
   \hline
   \cite{Surv-Conf028} & 2020 & no & no & no \\
    \hline
   \cite{Surv-Conf029} & 2020 & no & no & no \\
   \hline
   \cite{Surv-Conf030} & 2015 & no & no & no \\
    \hline
   \cite{Surv-Conf031} & 2015 & no & no & no \\
    \hline
   \cite{Surv-Conf032} & 2021 & no & no & no \\
    \hline
   \cite{Surv-Conf033} & 2015 & no & no & no \\
   \hline
   \cite{Surv-Conf034} & 2015 & no & no & no\\
   \hline
   \cite{Surv-Conf035} & 2015 & no & no & no\\
   \hline
   \cite{Surv-Conf050} & 2015 & no & no & no\\
   \hline
   \cite{Surv-Conf051} & 2017 & no & no & no\\
   \hline
   \cite{Surv-Conf052} & 2018 & no & no & no\\
   \hline
   \cite{Surv-Conf053} & 2019 & no & yes & no\\
   \hline
   \cite{Surv-Conf054}& 2019 & no & no & yes\\
  \hline
   \cite{Surv-Conf055} & 2018 & no & no & no \\
    \hline
   \cite{Surv-Conf056} & 2021 & no & no & yes \\
    \hline
   \cite{Surv-Conf057} & 2021 & no & no & yes \\
    \hline
   \cite{Surv-Conf058} & 2020 & no & no & yes \\
    \hline
   \cite{Surv-Earl-Jour002-1} & 2016 & no & no & yes \\
    \hline
   \cite{Surv-Earl-Jour005} & 2021 & no & no & yes \\
    \hline
   \cite{TYBH-Conf001} & 2020 & no & no & yes \\
     \hline
  \hline
  \end{tabular}
  \end{center}
\end{table}

\section{Ecosystem}

Privacy-Preserving Cloud can have many uses in different sectors in order to create an ecosystem \cite{b1}. In Section 3.1, we explore the various applications of this technology and examine how it can be beneficial in these areas. The technology which allows for these applications to function is explored in Section 3.2. In order to keep these ecosystems secure, there are various technologies and mechanisms which can be used. Section 3.3 covers several of these security measures. Finally, in Section 3.4, this paper will take a look at the security challenges which come along with using a cloud-based ecosystem.

\begin{figure}[h]
	    \centering
	    \includegraphics[width=0.9\linewidth,keepaspectratio]{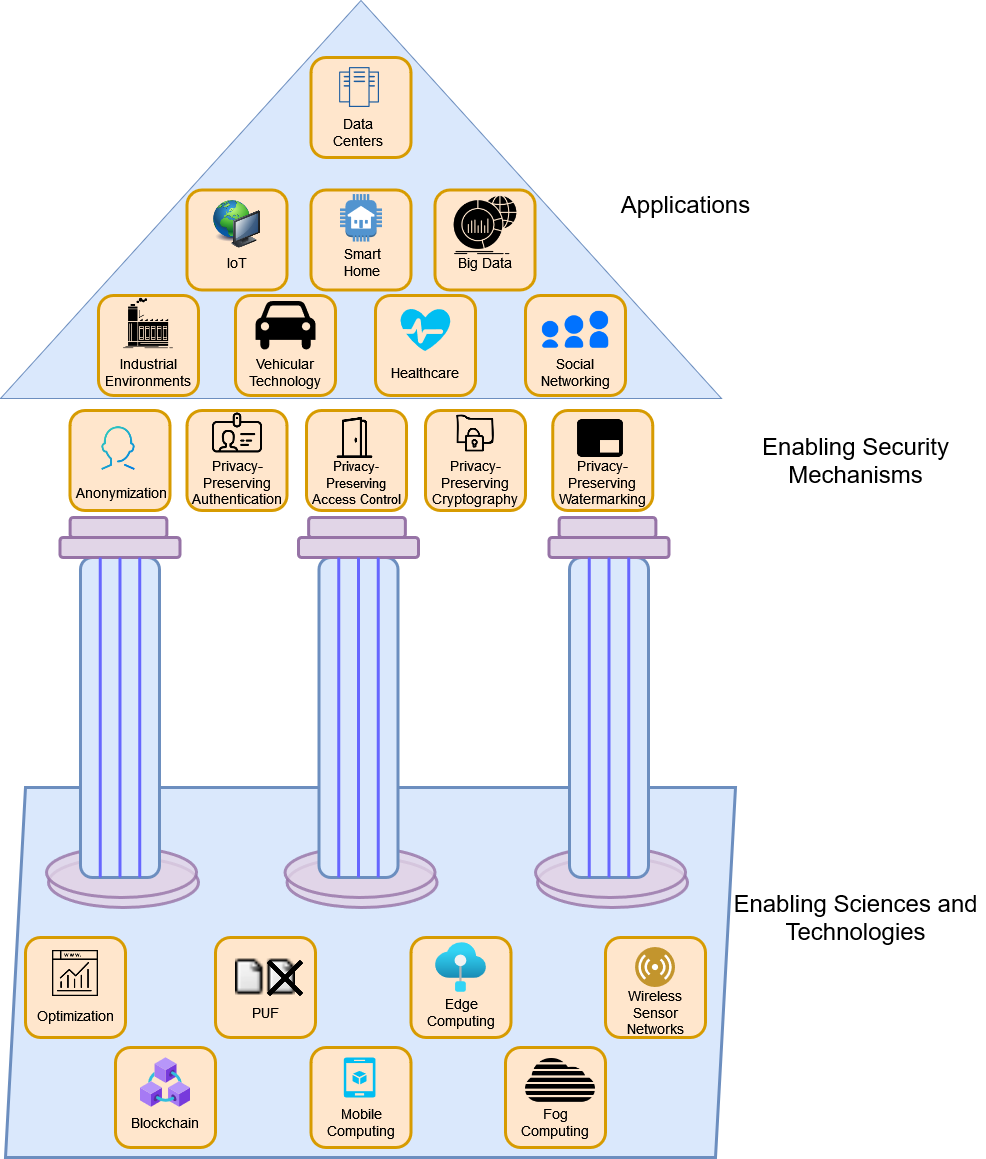}
	    \textcolor{black}{
	    \caption{Applications, Enabling Security Mechanisms, and Enabling Sciences and Technologies joining to form an Ecosystem for Privacy-Preserving Cloud.}\label{MyFig3}}
\end{figure}

\textcolor{black}{ In Figure \ref{MyFig3}, a base for the ecosystem is formed by Enabling Sciences and Technologies such as Optimization \cite{b2}, Blockchain \cite{b3}, PUF, Edge Computing, Fog Computing, and Wireless Sensor Networks \cite{b4}. On top of this base stands the pillars consisting of the Enabling Security Mechanism such as Anonymization, Privacy-Preserving Authentication, Privacy-Preserving Access Control, Privacy-Preserving Cryptography, and Privacy-Preserving Watermarking. Finally, Applications rest on top of the structure because they are enabled by the technologies below them. These applications include Data Centers, IoT, Smart Homes, Big Data, Industrial Environments, Vehicular Technology, Healthcare, and Social Networking.}

\subsection{Applications}

\subsubsection{Industrial Environments and IIoT}
The Industrial Internet of Things is a subset of the Internet of Things which works to optimize industrial production by providing tools with additional connectivity and intelligence, as detailed in \cite{State-Non-Earl-Jour001,b5}. In the healthcare industry, the Internet of Things works to improve services for patients with devices that can more effectively monitor them and collect their health data. \cite{State-Non-Earl-Jour002} describes how devices implanted in or worn by a patient can allow for a doctor to remotely diagnose a patient. \cite{State-Non-Earl-Jour003} outlines a method of protecting the privacy of industrial data is to encrypt it before sending it to the cloud, have operations performed on the encrypted data, and returning encrypted results to be decrypted by the user.

\subsubsection{Vehicular Technology}
\textcolor{black}{The ourline of \cite{State-Non-Earl-Jour004} indicates} how cloud technology integrated into vehicles allows for road traffic to be improved with devices such as sensors, cameras, and wireless transceivers.  The authors of \cite{State-Non-Earl-Jour005} say that it also helps create a network for autonomous vehicles so that data can be stored and processed remotely.\textcolor{black}{In \cite{State-Non-Earl-Jour006}, the authors proposed} the use of this technology in vehicular advertisement dissemination which allows for vehicles to receive advertisements from a cloud network . Vehicular cloud technology allows for more complex computations, such as calculating traffic and determining optimal routes, to be offloaded to parked cars which do not have to perform computations of higher priority, as said by \cite{State-Non-Earl-Jour007}.

\subsubsection{Healthcare}

\textcolor{black}{In \cite{State-Earl-Jour001}, authors state} that cloud-based data-driven healthcare monitoring services have allowed for a rapid growth of healthcare data which assist with eHealthcare. To protect this medical data, some networks will use a blind cloud framework which replaces patient identities with generated pseudo-identities. The authors of \cite{State-Non-Earl-Jour008} describe how this allows for unidentified data to be stored and analyzed. Searchable encryption schemes can also be used so that health data can be searched for while maintaining privacy as outlined by \cite{State-Non-Earl-Jour009}.  \textcolor{black}{Searchable symmetric encryption (SSE) is a method that provides the ability for users to utilize the storage of another party (a server) for privately hosting data that is encrypted. Therefore, the cloud service provider (CSP) does not have access to them, and when a user wants to access the data, CSP provides the encrypted data for them. }

\subsubsection{Social Networking}

In \cite{State-Non-Earl-Jour010}, it is described how many social networking platforms store user information on cloud services, so it is very important to keep this data secure. Techniques such as homomorphic encryption, oblivious transfer, and secret sharing can be used to protect user data. \textcolor{black}{"Homomorphic encryption is a method with which users can analyze or manipulate their encrypted data without decryption.}

\subsubsection{IoT}
Cloud technology has improved the Internet of Things’ ability to manage large amounts of data. Flexible privacy-preserving data sharing allows a user to send encrypted data using an identity-based encryption scheme that is outlined in \cite{State-Non-Earl-Jour011}. In addition, a network attestation scheme outlined by \cite{State-Non-Earl-Jour012} can be used to ensure that scalability, forward-security, and privacy.

\subsubsection{Smart Home}
Privacy-preserving smart homes have the ability to connect a single house to a cloud platform which aggregates the data from all other houses on the network. \textcolor{black}{The ourline of \cite{State-Non-Earl-Jour014} indicates} that this data can be used to increase home safety and energy efficiency in order to increase a residents quality of life.

\subsubsection{Big Data}
Because of mobile networks, data is constantly being generated at a rapid pace. \textcolor{black}{ In \cite{State-Non-Earl-Jour015}, authors state} that this data generation is called big data stream. The authors of \cite{State-Non-Earl-Jour016} outline how how cloud computing can offer a scalable infrastructure so that this big data stream can be processed and aggregated.

\subsubsection{Data Centers}
As detailed by the authors of\cite{State-Non-Earl-Jour017}, Geo-distributed cloud data centers hold and transfer large amounts of data every day. Due to privacy laws differing based on region, it can make the transfer of data across borders difficult to do. Data centers must ensure that they follow the strict data regulations put in place by countries and organizations, such as \textcolor{black}{ General Data Protection Regulation (GDPR)}.

\subsection{Enabling Sciences and Technologies}

\subsubsection{Optimization}
\textcolor{black}{In \cite{State-Earl-Jour002} , the authors describe} how outsourcing computation-intensive tasks to the cloud allows for systems to be more optimized, but comes at the risk of more privacy and security issues.

\subsubsection{Blockchain}
In order to protect data stored on the Cloud, blockchain-based technology \cite{c8} such as AuthPrivacyChain can be used to prevent hackers from accessing resources as outlined in the framework devised by \cite{State-Non-Earl-Jour018,b6}. \textcolor{black}{In the \cite{State-Non-Earl-Jour019} paper, it describes} how blockchain has also been used to implement forward and backward privacy by enforcing searchable public-key encryption, and \cite{State-Non-Earl-Jour020} describes how it can act as a distributed public ledger so that health data can be shared on a cloud-based system.

\subsubsection{PUF}

In \cite{State-Non-Earl-Jour021}, it is said that physical unclonable functions, or PUFs, provide the ability to have lightweight physical identities for smart devices.\textcolor{black}{Physical Unclonable Functions (PUFs) }introduce an extra player of physical security without bringing any additional overhead to smart devices.

\subsubsection{Mobile Computing}
Mobile cloud computing allows for users to access cloud services through wireless networks \cite{b7}. However, \cite{State-Non-Earl-Jour022} outlines how wireless networks are vulnerable to numerous threats which means more security measures must be taken. Some such measures which can be taken are listed by \cite{State-Non-Earl-Jour023}, such as key management for unlinkability, a secure indexing method, and integration of attribute based encryption. In addition, the authors of \cite{State-Non-Earl-Jour024} outline how scalable and privacy preserving friend matching offers a scalable solution for protecting social media user data over mobile networks. Authentication methods such as privacy-aware authentication and MediBchain-based privacy-preserving mutual authentication can also help protect data over mobile cloud networks, as described by the authors of \cite{State-Non-Earl-Jour025} and \cite{State-Non-Earl-Jour026}.

\subsubsection{Edge Computing}
In \cite{State-Non-Earl-Jour027}, it is described how mobile edge clouds deploy at the network edge in order to allow mobile users to access cloud services. However, mobile social sensing means that users need to report personal information which could result in data theft. \textcolor{black}{In \cite{State-Non-Earl-Jour028}, the authors propose} that participant grouping can work as one solution by leveraging secure sharing within groups.

\subsubsection{Fog Computing}
The authors of \cite{State-Non-Earl-Jour030} claim that fog computing can be used to extend cloud computations to the edge of a network and provide storage, geo-distribution, and mobile support. Fog computing can be used to keep data on the cloud safe by using a one-round authenticated key argument protocol as outlined by \cite{State-Non-Earl-Jour029}, or by using a homomorphic Paillier cryptosystem as outlined by \cite{State-Non-Earl-Jour031}.  \textcolor{black}{Homomorphic encryption is a method of encryption in which users can do mathematical operations on encrypted data without decryption.}
\textcolor{black}{ The Paillier cryptosystem, that Pascal Paillier has proposed in 1999, is a probabilistic asymmetric algorithm for public key cryptography.}

\subsubsection{Wireless Sensor Networks}
Wireless sensor networks are vulnerable to attack if the source node is located. \textcolor{black}{ "Wireless sensor network refers to wireless sensors that communicate and collect data, for instance, temperature, humidity, etc., from the environment that have been installed.}  \textcolor{black}{ In \cite{State-Non-Earl-Jour032}, authors proposed} that one method of protecting the location of the source node is to change packet destinations randomly in each transmission on the network. To compliment this, a source-location privacy protection scheme based on anonymity cloud which can also be used to protect to the location of the source node by constructing an anonymity cloud around the source node  \textcolor{black}{ has been proposed in \cite{State-Non-Earl-Jour033}.}

\subsection{Enabling Security Mechanisms}

\subsubsection{Anonymization}
In order to check that cloud infrastructure is meeting the intended security standards, the practise of security auditing is often employed \cite{b8}. In order to maintain privacy, input data and audit results are anonymized.\textcolor{black}{"Data Anonymization is the process that removes or alters the data that causes a subject to be identified directly.}  a method called SegGuard uses per-tenant encryption and property-preserving encryption so that the data is able to be audited while still offering sufficient protection \textcolor{black}{ has been proposed in \cite{State-Non-Earl-Jour034}.}

\subsubsection{Privacy-Preserving Authentication}
Many authentication schemes for tamper-proof devices have been proposed to allow privacy and security in open-access environments. \textcolor{black}{In \cite{State-Earl-Jour003} , the authors explain} how Vehicular Ad-Hoc Networks require such authentication so that they may improve transportation networks. Because of the cloud, these vehicles need to only register with a trusted authority once in order to authenticate as outlined by the authors of \cite{State-Non-Earl-Jour038}. \textcolor{black}{In \cite{State-Non-Earl-Jour035} the authors suggest }that cloud computing has allowed for data from biometric identification to be outsourced to the cloud for computation, offering a more reliable way to authenticate a user.\textcolor{black}{ The authors of \cite{State-Non-Earl-Jour037} }explain how this biometric data can be protected by using a matrix-transformation-based privacy-preserving biometric identification scheme using the property of the orthogonal matrix and additional randomness. Additionally, \textcolor{black}{the authors of \cite{State-Non-Earl-Jour036} detail} how authentication can also be improved for mobile users by using the cloud to support mutual authentication, key exchange, user anonymity, and user untraceablilty.\textcolor{black}{ In \cite{State-Non-Earl-Jour039}, the authors suggest }that a shared authority based privacy-preserving authentication protocol can be used to address the issue that a challenged privacy request on a shared network can reveal a user’s privacy regardless of whether their request is granted. In a distributed cloud-based system, an authorized accessible privacy model can be used to keep data and identity private simultaneously \textcolor{black}{ has been presented in \cite{State-Non-Earl-Jour040}.} Anonymous authentication can be achieved by using a registered security token in conjunction with a key exchange which uses preauthentication and postauthentication user anonymity to generate a pseudoidentity \cite{State-Non-Earl-Jour041}.

\subsubsection{Privacy-Preserving Access Control}

In order to facilitate resource sharing, the authors of \cite{State-Non-Earl-Jour044} suggest that cloud applications allow users to delegate their access permissions. Access control policies guard the access to cloud resources and credentials. In the bioinformatic field, large amounts of genomic data are stored and processed over the cloud. The P2GT scheme detailed by \cite{State-Earl-Jour004} uses key-policy attribute-based encryption to manage data access control. Mobile cloud computing can use authentication schemes such as \textcolor{black}{Self-Certified Public Key Cryptography (SCPKC), Chinese Remainder Theorem (CRT), Ciphertext-Policy Attribute-Based Encryption (CP-ABE), Scalable Media Access Control and Deduplication and (SMACD)} to manage access, as detailed by the authors of \cite{State-Earl-Jour005}, \cite{State-Non-Earl-Jour042}, and \cite{State-Non-Earl-Jour045}. In order to enforce authorization for fine-grained access control on the cloud,\textcolor{black}{ the authors in \cite{State-Non-Earl-Jour043} state }that an organization must use encryption.

\subsubsection{Privacy-Preserving Cryptography}

\paragraph{Data Encryption}\mbox{}

According to \cite{State-Earl-Jour046}, users can search encrypted documents on the cloud using dynamic symmetric searchable encryption. Dynamic data encryption allows for performance to be improved while working under time constraints \cite{State-Non-Earl-Jour047} and achieves a high level of privacy \cite{State-Earl-Jour006}.\textcolor{black}{ The authors in\cite{State-Non-Earl-Jour048} suggest }using certificateless encryption keyword search to allow for multirecipient keyword searching, which allows for privacy to be preserved on the Internet of Things. Because the cloud is so open, \textcolor{black}{ in \cite{State-Non-Earl-Jour049} the authors claim} that identity-based encryption is important to maintain accountability. The authors of \cite{State-Non-Earl-Jour051} state that user identities can even be formed by combining a user’s identity with the identity of their attribute authority. Using asymmetric scalar-product-preserving encryption, \textcolor{black}{in \cite{State-Non-Earl-Jour050}  authors indicate that} fuzzy multi-keyword search can be achieved.

\paragraph{Encrypted Data Processing}\mbox{}
With the popularity of cloud computing, \textcolor{black}{the authors of  \cite{State-Earl-Jour007} claim }that more and more documents are encrypted before they are outsourced. Due to concerns about the privacy of data on the cloud,\textcolor{black}{ the authors in \cite{State-Earl-Jour008} suggest }a geometric range query scheme can be used to protect the location of the data . A keyword search is a common way of filtering encrypted data for use, according to \cite{State-Non-Earl-Jour052}, and can be performed using \textcolor{black}{ Multi-keyword Ranked Search over Encrypted (MRSE) and Multi-keyword Ranked Search over Encrypted data in Hybrid Clouds (MRSE-HC) schemes}, as per \cite{State-Non-Earl-Jour053} and \cite{State-Non-Earl-Jour054}. To verify the results of these multi-keyword searches, a scheme known as VPSearch \textcolor{black}{ has been introduced in \cite{State-Non-Earl-Jour058}} which can be used. Another way to efficiently access data is to use a privacy-preserving outsourced similarity test in order to capture the desired data which has been laid out by the authors of \cite{State-Non-Earl-Jour055}. Frequent itemset mining is a popular way to data mine on large cloud datasets \cite{State-Non-Earl-Jour059}. To analyze big graphs that have been outsourced to the cloud, spectral analysis can be used as desecribed by \cite{State-Non-Earl-Jour056}. When querying for graphs, constrained shortest distance can help by finding the shortest distance from an origin to a destination of a graph \cite{State-Non-Earl-Jour060}. Cloud-based Personal Health Record systems are used to manage health records securely by ensuring that submitted data is encrypted \cite{State-Non-Earl-Jour057}. Data retrieval confidentiality can be supported using Searchable Encryption, as outlined by \cite{State-Non-Earl-Jour061}, to allow a trusted server to access user data. \cite{State-Non-Earl-Jour062} says that tensor decomposition can also be used so that multiple users can have their data retrieved without the cloud learning about the users data.

\subsubsection{Privacy-Preserving Watermarking}
\textcolor{black}{In \cite{State-Non-Earl-Jour063}, authors define } Watermarking as when a watermark is embedded into media so that it may be identified. Privacy-Preserving Watermarking allows for a party to entrust the task of watermarking data to a cloud service.

\subsubsection{Privacy-Preserving Integrity Checking and Auditing}
While cloud storage offers easy maintenance and management of stored data, \textcolor{black}{the authors of \cite{State-Non-Earl-Jour064} say} that it does not ensure integrity. Remote data integrity checking can be used to verify that data stored on the cloud is stored honestly \cite{State-Non-Earl-Jour065}. A third party auditor can be used to audit data to ensure integrity, as can dynamic data integrity auditing as stated by the authors of \cite{State-Earl-Jour009} and \cite{State-Non-Earl-Jour066}. A Provable Data Possession proposed by \cite{State-Non-Earl-Jour067} can allow people to audit data efficiently by outsourcing data into small blocks of data and then blinding them so that each block can be individually verified.

\subsection{Related Security Challenges}

\subsubsection{Trust and Reputation}

\paragraph{Trustable Computing}\mbox{}
The authors of \cite{State-Non-Earl-Jour068} state that in Cloud Computing, privacy and trust create an inherent conflict. In order to evaluate the quantification of privacy and trust in a cloud system,\textcolor{black}{ in \cite{State-Non-Earl-Jour069}, the authors state }that both must be weighted accordingly. A Reputation-aware Trust and Privacy scheme proposed by \cite{State-Non-Earl-Jour070} can deal with trust management by leveraging trust scores for cloud services.

\paragraph{Computation in Untrusted Environments}\mbox{}
Since outsourcing spatial keyword searches to an untrusted cloud can cause privacy concerns,\textcolor{black}{ in \cite{State-Non-Earl-Jour071}, the authors proposed} that top-k spatial keyword searches can be performed instead, where spatial and textual data are encrypted in a unified way. When an untrusted cloud environment is to be used, a privacy-preserving framework using homomorphic cryptography outlined by \cite{State-Non-Earl-Jour072} can be used to conduct the needed computations.

\subsubsection{Attack Resilience}
Wireless sensor networks are often deployed in open and large areas which makes them vulnerable to attack. To protect the location of the source node, \textcolor{black}{the authors of \cite{State-Non-Earl-Jour073} describe} how fake traffic can be created to camouflage the source node in the cloud. If the guest operating system or the virtual machine monitor of a cloud vendor are compromised, an attack can be launched against users. A solution to this problem put forward by \cite{State-Non-Earl-Jour074} is to integrate on-chip memory protection for trusted applications from attacks.

\subsubsection{Forensics}
The aim of cloud forensics is to process large sets of data from the cloud and recover evidence. A multigranularity privacy leakage forensics method put forward by the authors of \cite{State-Non-Earl-Jour075} allows for analysis of privacy violations which have been caused by malware.

\section{Life Cycle}

The Life Cycle of a Privacy-Preserving Cloud system can be broken down into four stages. The first subsection covers Design, which outlines what factors must be taken into account when designing a system. Verification is covered in the second subsection and Implementation is covered in the third. The fourth and final subsection looks at how a system of this nature would be deployed and evaluated.

\begin{figure}[h]
	    \centering
	    \includegraphics[width=1\linewidth,keepaspectratio]{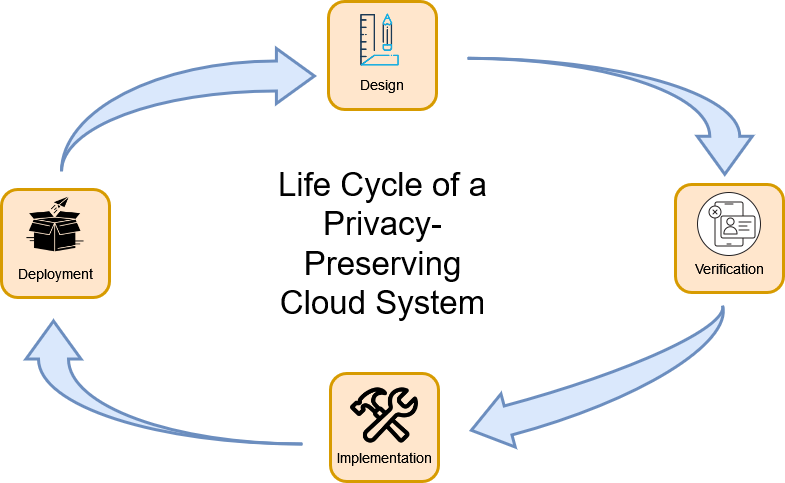}
	    \textcolor{black}{
	    \caption{Design, Verification, Implementation, and Deployment form the Life Cycle of a Privacy-Preserving Cloud System.}\label{MyFig4}}
\end{figure}

\textcolor{black}{ In Figure \ref{MyFig4}, the formation of a Life Cycle of a Privacy-Preserving Cloud System is illustrated, with each of the four steps in sequence. The arrows connecting each box signifies the order in which the steps occur. }

\subsection{Design}

\subsubsection{Additional Design Objectives}

\paragraph{Cost-Effectivity}\mbox{}

Because encrypting all intermediate data sets stored on the cloud would not be cost-effective, \textcolor{black}{ the authors in \cite{State-Non-Earl-Jour076} suggest }that it is more efficient to first identify which data sets require encrypting so that data privacy is maintained while keeping costs low.

\paragraph{Efficiency}\mbox{}
In \cite{State-Non-Earl-Jour077}, it is said that location-based services found in wireless communications present new efficient schemes in outsourced cloud where the users identity is masked for privacy purposes. By designing a data sharing process which provides integrity by design, \textcolor{black}{in \cite{State-Non-Earl-Jour078} ,authors state }that efficiency can be improved by have a third party audit data stored on the cloud.

\paragraph{Resource-Constraint Awareness}\mbox{}
Vehicular cloud computing consists of many computations that must be performed dynamically \cite{b9}, but the limited computational capabilities of a single vehicle must be taken into account. To create a functioning system, \textcolor{black}{the authors of \cite{State-Non-Earl-Jour079} insist }that a lightweight architecture must be designed. Likewise, in the healthcare system, heavy computations must be performed on limited hardware, necessitating the use of a lightweight resource-aware system, such as the one proposed by \cite{State-Non-Earl-Jour080}. A lightweight Delegatable Proofs of Storage scheme proposed by the authors of \cite{State-Non-Earl-Jour081} can also be used to speed up data integrity checks in cloud storage systems.

\paragraph{Time Constraint Awareness}\mbox{}
Because cloud storage can contain a large number of images, it is important for security and privacy reasons that data is audited as soon as possible. For this reason, \textcolor{black}{the authors of \cite{State-Non-Earl-Jour082}, state }that it is important that cloud systems employ auditing schemes that are aware of time constraints.

\paragraph{Reliability}\mbox{}
Cloud storage services have become a very popular way to store data, but there are risks of data loss and data leaks. \textcolor{black}{The authors of \cite{State-Non-Earl-Jour083} demand} that these issues must be addressed in order for a cloud storage service to be reliable.

\paragraph{Scalability}\mbox{}

Because cloud services are used to support big data applications in sectors such as business and healthcare, it is important that they have a scalable infrastructure. A proximity privacy model proposed by \cite{State-Non-Earl-Jour084} can be used to cluster data so that high scalability can be achieved by performing data-parallel computations.

\paragraph{Performance}
One way to improve the performance of a cloud system is to exploit geo-distributed clouds to reduce service delay in a method proposed by the authors of \cite{State-Non-Earl-Jour085}. Algorithms such as PBtree traversal width minimization and PBtree traversal depth minimization can also be used to increase performance by increasing the efficiency of query processing when used as outlined by \cite{State-Non-Earl-Jour086}.

\subsection{Verification}
In order to address the risk of private data leaking to the public, a Structural Integrity Verification method described by \cite{State-Non-Earl-Jour087} can be applied to cloud systems using three processes: proof organization, proof transformation, and integrity judgement. For large sets of mobile data, the batch verification method proposed by \cite{State-Non-Earl-Jour088} can be used to reduce the computational cost.

\subsection{Implementation}

\subsection{Deployment}
When an organization chooses to use a cloud service, they must consider the security and privacy risks which they are willing to take. For this reason, \textcolor{black}{the authors of \cite{State-Non-Earl-Jour089} state} that it is important that a cloud service based on an organizations needs in these areas.

\subsubsection{Evaluation}

Using advancements in deep learning technology, the authors of \cite{State-Non-Earl-Jour090} state that attacks against cloud systems can be analyzed in order to propose new defense methodologies.\textcolor{black}{ In \cite{State-Non-Earl-Jour091}, authors emphasize }that it is important that quantitative assessment of the privacy practises of cloud services are analyzed so that clients can make an informed decision about which provided they choose to do business with.

\section{Layered Architecture}

he architecture of a cloud system is layered, with each layer having its own purpose. The first subsection examines the infrastructure of a cloud system which maintains privacy. The second subsection looks at how privacy is maintained with services such as pre-diagnosis computing and location-based services. The third section details how tasks can be managed privately using task allocation and computation offloading.

\subsection{Privacy-Preserving Infrastructure}

\subsubsection{Privacy-Preserving Physical Layer}

The authors of \cite{State-Non-Earl-Jour092} state that in order to audit cloud data to check for data integrity, two lightweight auditing protocols can be used based on online and offline signatures. A device needs only to perform a computation when an outsourced file is available.

\subsubsection{Privacy-Preserving Data and Data Infrastructure}

\paragraph{Privacy-Preserving Data}\mbox{}
As defined by \cite{State-Non-Earl-Jour093}, Data Deduplication is a technique which eliminates duplicate data stored on the cloud to save space and bandwidth. However, an attacker can abuse a deduplication check to determine if a file is already stored in a cloud system. Restricting deduplication checks to only the agent and the cloud service provider, data privacy can be improved, according to \cite{TBT-Jour001}.

\paragraph{Privacy-Preserving Storage Storage}\mbox{}
As outlined in \cite{State-Non-Earl-Jour094}, cloud storage systems allow for data to be outsourced flexibly and conveniently. The authors of \cite{State-Non-Earl-Jour095} detail how deduplication allows for cloud storage to only need to keep one copy of a given file to reduce space usage. Because data such as power meter readings require privacy, schemes such as privacy-friendly cloud storage as detailed in \cite{State-Non-Earl-Jour096} are required. Privacy and reliability in cloud storage environments can be achieved using encryption schemes proposed by the authors of \cite{State-Non-Earl-Jour097} such as BP-XOR secret sharing or LDPC secret sharing. Another way to preserve privacy proposed by \cite{State-Non-Earl-Jour098} is to enable cloud users to distribute and search encrypted cloud data across clouds managed by CSPs. To safely share data, \textcolor{black}{the authors in \cite{State-Non-Earl-Jour099}, say }that sanitizable signatures can be used to hide sensitive information. Its is important that auditing is available in cloud storage systems so that users can have the integrity of their data ensured, as per \cite{State-Non-Earl-Jour100}. In order to properly audit regenerating-based cloud storage, a proxy that is privileged to regenerate authenticators must be used in a way detailed in \cite{State-Non-Earl-Jour101}. To protect cloud servers from internal attacks, a three-layer storage framework devised by \cite{State-Non-Earl-Jour102} can be used based on fog computing.

\paragraph{Privacy-Preserving Databases}\mbox{}

In \cite{TBT-Jour002}, it is outlined why multi-dimensional range queries on cloud networks should not be fully trusted as many implementations suffer from data leakage. A blockchain enabled query framework can result in high efficiency and scalability \cite{State-Earl-Jour010}. For health data, a query scheme put forth by \cite{State-Non-Earl-Jour103} could be used where medical data is encrypted and stored on the cloud, and then specific diseases are queried on that data. Searchable symmetric encryption, as proposed by \cite{State-Non-Earl-Jour104}, can be used to protect the privacy of both the queries and the database. For a secure database, a two-cloud architecture devised by \cite{State-Non-Earl-Jour105} can also be used, where a series of intersection protocols provide privacy for numeric-related range queries.

\subsubsection{Privacy-Preserving Data Exchange Infrastructure}

\paragraph{Privacy-Preserving Data Sharing}\mbox{}

Data sharing is a service that, when supplied by cloud computing, the authors of \cite{State-Non-Earl-Jour106} say can be convenient and economic to a user. \textcolor{black}{In \cite{State-Non-Earl-Jour107}, authors emphasize }that it is important that the data being shared is not exposed to unauthorized users or to the cloud provider in order to maintain privacy, which generally comes with a computational cost as said in \cite{State-Earl-Jour011}. Secret sharing group key management protocol and public auditing can be used to help tighten security when data sharing in methods detailed by the authors of \cite{State-Non-Earl-Jour108} and \cite{State-Earl-Jour012}. Ring signatures can help to compute verification metadata when performing a data audit when used as described by \cite{State-Non-Earl-Jour109}, and a broadcast group key management system detailed by \cite{State-Non-Earl-Jour110} can help with selectively sharing documents.

\paragraph{Privacy-Preserving Data Publishing}\mbox{}
With cloud computing, because data owners are no longer in possession of physical copies of their own data, this creates the issue of privacy-preserving set-valued data publishing as said in \cite{State-Non-Earl-Jour111}. Using retrievable data perturbation as detailed by the authors of \cite{State-Non-Earl-Jour112}, we can create a solution to this issue.

\subsubsection{Privacy-Preservig Communications}
To deal with the growing amount of network traffic, some traffic is redirected to outsourced middleboxes. To ensure data privacy during redirection, encrypted header-matching schemes proposed by \cite{State-Non-Earl-Jour113} can be used to keep data safe. Network security can also be enhanced by protecting transmitted social messages with integrated security-based mobility prediction algorithms to help resist attacks, as detailed in \cite{State-Non-Earl-Jour114}.

\subsubsection{Privacy-Preserving Protocols}
When ensuring security during data transmission and storage, there should be protocols set in place as describer in \cite{State-Non-Earl-Jour115} to protect the identify of communicating units during encryption. To evaluate a protocol, \textcolor{black}{the author in \cite{State-Non-Earl-Jour116} state }that a simulation can be performed based on typical circumstances of attacks.

\subsection{Privacy-Preserving Services}
In the mobile healthcare field, a Diverse Keyword Searchable Encryption scheme proposed by \cite{State-Earl-Jour013} can be used to search and make diagnosis over stored encrypted data. To maintain privacy of genomic data, the authors of \cite{State-Earl-Jour014} state that information should only be access through a fine-grained access control policy with attribute-based encryption. To preserve the privacy of location-based services, data outsourced to the cloud should be encrypted and only be able to be queried by registered authorized users in methods detailed by the authors of \cite{State-Earl-Jour015} and \cite{State-Non-Earl-Jour117}. For data sharing services, \textcolor{black}{the authors of \cite{State-Earl-Jour016} state } that data should not be shared with a third-party in order to maintain privacy.

\subsection{Privacy-Preserving Computing}

\subsubsection{Privacy-Preserving Task Management}

\paragraph{Privacy-Preserving Task allocation}\mbox{}
In a mobile network, an auction-based privacy-preserving incentive scheme proposed in \cite{State-Non-Earl-Jour118} can be used for task allocation, where each user can control their privacy budget. Using differential privacy and geocast, we can create task allocation strategies that consider privacy, utility, and overhead in a mobile environment as detailed in \cite{State-Non-Earl-Jour119}.

\paragraph{Privacy-Preserving computation Offloading}\mbox{}
Computation offloading allows for devices to offload tasks to the cloud in order to preserve energy and time. When offloading, usually devices will do so to the nearest edge cloud which can expose the location of the device. This issue can be resolved with a location privacy-guaranteed offloading algorithm proposed in \cite{State-Earl-Jour017}.

\subsubsection{Privacy-Preserving Task}

\paragraph{Privacy-Preserving Search}\mbox{}

\begin{itemize}

\item Keyword Search

Many approaches towards searching over encrypted data are only able to perform exact searches, as per \cite{State-Non-Earl-Jour120}. A privacy preserving ranked multi-keyword search in a multi-owner model as proposed in \cite{State-Non-Earl-Jour121} allows for a user to search without knowing keyword data. With multi-keyword search, a tree-based structure index can improve efficiency, as outlined in \cite{State-Non-Earl-Jour122} and \cite{State-Non-Earl-Jour123}.

\item Content Search

\end{itemize}

A privacy-preserving image search system on cloud as outlined in \cite{State-Non-Earl-Jour124} would help protect users privacy by guarding the sensitive information that is stored in image files such as location and event.

\paragraph{Privacy-Preserving Data Analytics}\mbox{}
Privacy-preserving data classification techniques such as the ones in \cite{State-Non-Earl-Jour125} can be used where knowledge about client input data is not made available to the cloud server. In order to provide for the needs of elderly people in need of housing, \textcolor{black}{in  \cite{State-Non-Earl-Jour126} the authors say }that real estate recommendations can be made based on the behaviour of the elderly in the cloud. By analyzing sports records, \cite{State-Non-Earl-Jour127} says that exercise recommendations can be made for individuals who wish to live healthier. Likewise, a clinical cloud system could use data to make diagnoses, as outlined in \cite{State-Non-Earl-Jour130}. Since the client cannot directly access this data, decision tree training and k-nearest neighbor computation can be used as proposed in \cite{State-Non-Earl-Jour128} and \cite{State-Non-Earl-Jour129}.

\paragraph{Privacy-Preserving Content Processing}\mbox{}

\begin{itemize}

\item Image

\begin{itemize}

\item Image Retrieval

Cloud computing has allowed for large amounts of images to be stored easily as outlined in \cite{State-Earl-Jour018}, but this has led to a number of privacy concerns, as per \cite{State-Non-Earl-Jour131}. In \cite{State-Non-Earl-Jour132}, it is described how content-based image retrieval applications have generally been limited by computation and storage requirements. A polyalphabetic cipher can be used to improve security with no retrieval accuracy degradation, according to \cite{State-Non-Earl-Jour133}, and k-ranked multi-keyword search defend against keyword guessing attacks when used as outlined in \cite{State-Non-Earl-Jour134}.\textcolor{black}{ In \cite{State-Non-Earl-Jour135}, the authors outline} how watermark protocols can also be used to prevent from unauthorized image copying.

\item Image Reconstruction
In \cite{State-Non-Earl-Jour136}, it is outlined how an outsourced image recovery service allows for users to only need to outsource compressed images to the cloud in order to reduce the amount of storage that they use. The cloud can then reconstruct the compressed images safely.

\item Image Denoising

\end{itemize}

While image denoising can help to restore an image, \cite{State-Non-Earl-Jour137} claims that it also raises security and privacy concerns. One such denoising method enable encrypted databases to provide query-based image denoising.

\item Video

\begin{itemize}

\item Video Surveillence
During video surveillance, it is important that privacy is protected without interfering with the regular tasks of the surveillance, say the authors of \cite{State-Non-Earl-Jour138}. The authors of \cite{State-Non-Earl-Jour139} describe how Privacy Region Protecting protects privacy regions without effecting the non-privacy regions so that they are visually intact.

\item Video Reporting

In \cite{State-Non-Earl-Jour140} it is described how in vehicular networks, real-time video reporting can instantly report traffic accidents for safety reasons. Using a 5G network, it can be ensured that the system is reliable and secure.

\end{itemize}

\end{itemize}

\paragraph{Privacy-Preserving Cloud Auditing}\mbox{}
Problems such as privacy disclosure, authority abuse, and collusion attacks can be solved by cloud auditing as detailed in \cite{b10,State-Non-Earl-Jour141}. One such scheme proposed in \cite{State-Non-Earl-Jour142} allows for user anonymity without group signatures so that tags are compact. As a real world example, unmanned aerial vehicles require an audit scheme to accommodate their limited storage and computing performance. They can use an audit scheme such as the one in \cite{State-Non-Earl-Jour143} using a distributed string equality check and Merkle-hash tree multi-level index structure.

\paragraph{Other Kinds of Privacy-Preserving Computing}\mbox{}

Cloud computing has been used in other such ways, as detailed in \cite{State-Non-Earl-Jour144}, such as Pockit, a calculation toolkit which outsources data to perform common arithmetic securely. \textcolor{black}{In \cite{State-Non-Earl-Jour145}, authors claim} that a secure and efficient privacy preserving provable data possession scheme can be used to allow auditors to verify data possession. Vehicular ad hoc networks allow vehicles to make real time computations and receive updates on things such as road conditions and accidents as outlined in \cite{State-Non-Earl-Jour146}. Finally, as described in \cite{State-Non-Earl-Jour147}, cloud systems have allowed for medical prediagnosis services to develop so that users can privately receive remote healthcare.

\begin{figure}[h]
	    \centering
	    \includegraphics[width=1\linewidth,keepaspectratio]{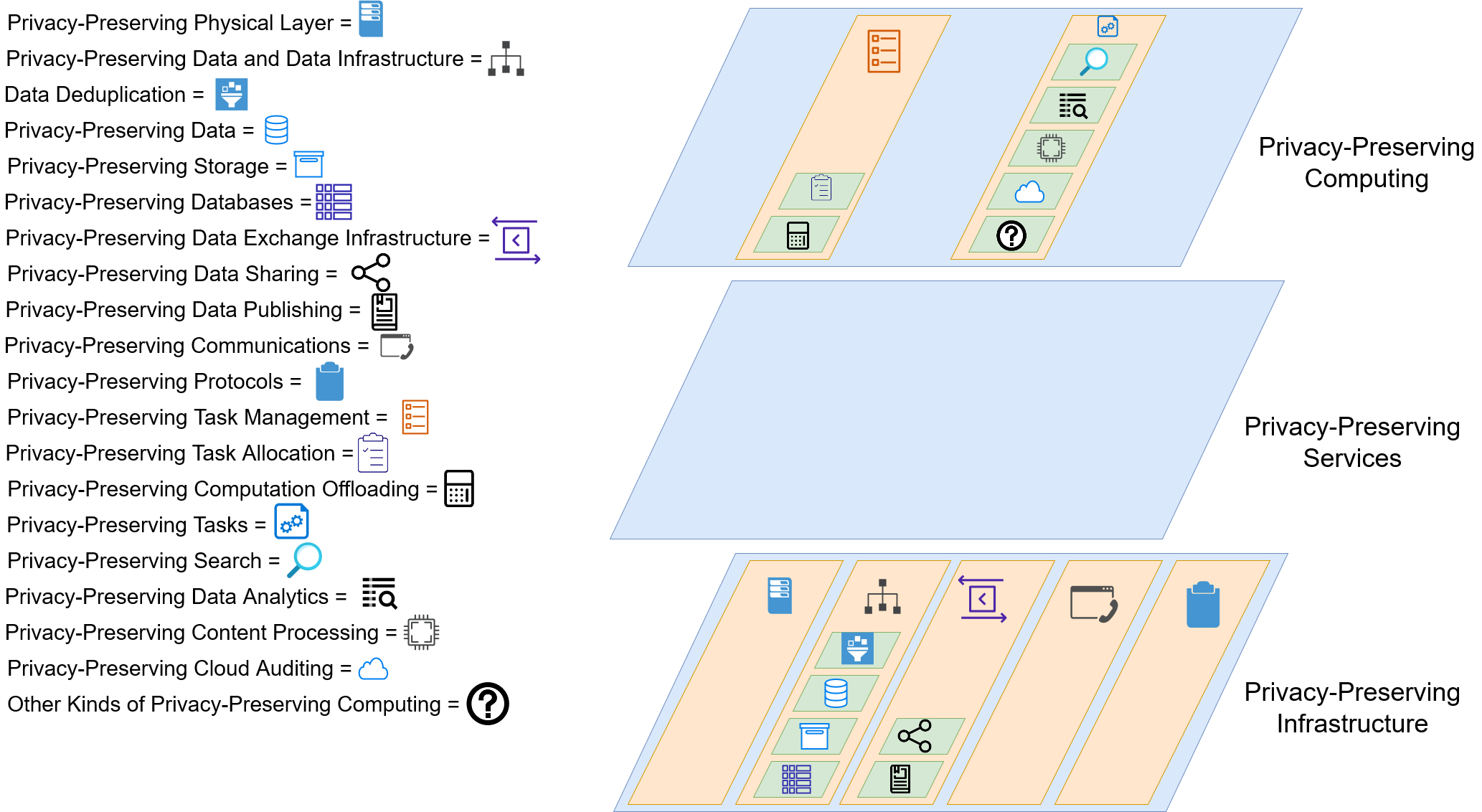}
	    \caption{The Layered Architecture of Privacy-Preserving Cloud}
	    \label{LayArch}
\end{figure}

\textcolor{black}{ In Figure \ref{LayArch}, the three large blue parallelograms represent the three main layers of the Layered Architecture. Within these layers are the sublayers represented by the orange parallelograms. Withing the sublayers are the subsublayers represented by the green parallelograms. On the left of the image is a legend which labels the symbols and what they represent.}

\section{Future Roadmap: the Promise of AI}\label{Fut}

We anticipate that research on privacy-preserving cloud will move towards quantum-inspired and bio-inspired AI-assisted privacy-preserving cloud in the near future. Our reason for such an anticipation is the existence of trends towards AI-assisted cloud, AI-assisted secure cloud and AI-assisted privacy-preserving cloud as well as quantum-inspired AI and bio-inspired AI, which are discussed in Subsections \ref{sub1}, \ref{sub2}, \ref{sub3}, \ref{sub4} and \ref{sub5}, respectively.

\subsection{AI in Cloud}\label{sub1}
AI models have been used in the vast amount of data the needs to be analyzed, these models can learn from the data that we have provided for training then works with a meager error rate \cite{c1}. \textcolor{black}{ Machine Learning and Probabilistic Analysis based Model (MLPAM)} is a novel model created with machine learning to preserve data in the Cloud environment. A comprehensive analysis indicates its security and efficiency in \cite{NGO-Jour001}, the authors have argued that MLPAM could provide safety and efficiency of data sharing and control in the future for multiple environments such as IoT while in \cite{Fut-Earl-Jour001}, An Automatic Cloud Detection neural network (ACD net) has been presented that contains two features which can overcome the overestimation problems in Cloud Computing. However, the authors have not created a future road map and explain that their solution cannot overcome the thin cloud detection issue because of an inadequate training dataset. In addition, the Asynchronous-Advantage-Actor-Critic (A3C) presented in \cite{Fut-Earl-Jour002} for decentralized learning to analyze the parameters of tasks and hosts to make a scheduling decision to perform better, as well as they introduced scalable reinforcement learning models that can be explored more in the future. Furthermore, for developing and managing Deep Neural Network (DNN) in the Cloud-to-Things Continuum, the authors in \cite{Fut-Earl-Jour003} have used Kafka-ML framework, an open-source framework used in Machine Learning and Artificial Intelligence management. Their analysis has indicated that response time has improved noticeably in comparison with Cloud-only deployment. Moreover, Since the mobile web has not had adequate resources to support Deep Neural Network (DNN) in cloud computing, the authors in \cite{Fut-Earl-Jour004} a Lightweight Collaborative Deep Neural Network called LcDNN has been presented in order to make use of Deep Neural Network (DNN) possible in the cloud for mobile. However, a new deep model named Multiscale Attention Convolutional Neural Network (MACNN) has been presented in In \cite{Fut-Earl-Jour005},  that used asymmetric encoder-decoder structure for ground-based cloud detection. In contrast, in \cite{Fut-Earl-Jour006}, a utility-based mechanism has been offered to equilibrium cost and availability in spot instances. Likewise, a Long Short Term Memory for Neural network framework has been used to examine price tendency. On the other hand, a method in neural network architecture named CloudU-Net for\textcolor{black}{ Ground-based Daytime and Nighttime Cloud Images (GDNCI)} used has been presented in \cite{Fut-Earl-Jour007}, that is used in the weather forecast; as observed in the paper, its performances have been improved for daytime and nighttime cloud images. As opposed to \cite{Fut-Earl-Jour008} that \textcolor{black}{An innovative Multilevel Attention-based U-shape Graph Neural Network (MAUGNN) }has been introduced, and the authors' purpose was using (MAUGNN) is to learn point cloud effectively due to 3-D sensors in the IoT industry. In addition, \textcolor{black}{in \cite{Fut-Earl-Jour009} a multiscale convolutional neural network with color vegetation indices (MCCNN) has been presented. The purpose of this model is labeling in point cloud for the 3-D model; also, the authors have analyzed its efficiency in Terrestrial Laser Scanning (TLS) data ( Semantic3D) and ALS data which is Vaihingen3D data point cloud.} Similar to \cite{Fut-Earl-Jour010} that \textcolor{black}{a Pipelined point cloud based Neural Network processor (PNN)} has been presented in order to low latency PNN-based 3-D intelligent systems in mobile devices. The authors in \cite{Fut-Earl-Jour010} have claimed the speed of algorithms has increased because of the 3-D points from the depth image sample and group directly. Furthermore, a novel spatiotemporal neural network with four modules has been introduced in order to remove cloud cover from satellite images in \cite{Fut-Earl-Jour011} not to mention its impact on the quality of remote sensing mapping and geoinformation has gotten better. As opposed to \cite{Fut-Earl-Jour012} that Neural Ordinary Differential Equations (NODEs) with Recurrent Neural Networks (RNNs) has been studied in order to assimilate image observations that are to say, with this method classifying missing data in cloud cover is possible. In \cite{Fut-Earl-Jour014} similarly, a new model was presented for detecting cloud, the method has been developed in Deep Neural Network (DNN) \cite{c4} technique with the cross-track infrared sounder, and authors have used the cloud information measured from the image to train their model. Moreover, a number of methods in the neural network, which can be in time-series dataset of cloud server power modeling, have been analyzed in \cite{Fut-Earl-Jour015}, also, the authors have studied the advantages and drawbacks of models and their effects. In contrast, deep neural networks have been presented in \cite{Fut-Earl-Jour016} to obtain deep-hidden fault information in cloud data centers by analyzing massive alarm data collected in their interconnections. In addition, due to storing and analyzing an illimitable amount of data is just possible in cloud computing, in \cite{Fut-Earl-Jour017}, a method for predicting the amount of data requisite cloud services has been presented that is a hierarchical Pythagorean fuzzy deep neural network. Furthermore, in order to improve the accuracy of occupancy map videos, the authors in \cite{Fut-Earl-Jour020} implement a Convolutional Neural Network (CNN). They claimed that this is the first learning-based work to improve the performance of \textcolor{black}{Videobased point Cloud Compression (V-PCC)}.

\subsection{AI in Secure Cloud}\label{sub2}

A federated-learning-enabled IoT scheme that provides security and efficiency has been presented in \cite{Fut-Earl-Jour021}, this framework has been produced in order to provide communication-efficient and privacy-preserving users' data in smart grids with Cloud collaborations similar to \cite{Fut-Earl-Jour022} that a secure, verifiable, and fair technique has been presented which protect the privacy of sensitive information, and it operates in an untrustworthy cloud-server. Furthermore, in order to preserve medical diagnosis from \textcolor{black}{Data Islands(DI)-level} poisoning attacks, a secure federated learning mechanism that guarantees the confidentiality of DI-related information named SFAP that make use of multiple keys possible has been presented in\cite{Fut-Earl-Jour023}. In contrast with \cite{Fut-Earl-Jour024} that authors have analyzed several neural network prediction services from the perspective of privacy-preserving, the privacy of the model, and the query are considered as well as examined several proposals and introduced an optimized neural network prediction scheme that brings about high accuracy, model privacy, and low overheads in the outsourcing setting. Moreover, In \cite{NGOR-Fut-Earl-Jour024}, different features have been created to symbolize each disability, then by these features, the image of people with a disability such as wheelchairs, blind people, and people with Down syndrome have been classified. The authors' result in \cite{NGOR-Fut-Earl-Jour024} has demonstrated that the level of securing image mobility in Cloud systems has been improved. In addition, the authors have introduced a framwork that provides training and testing data privacy in \cite{NBOR-Fut-Earl-Jour024}, this framwork is a secure cloud-intelligent network that is supported by privacy-preserving machine learning similar to \cite{NTOR-Fut-Earl-Jour024} that authors have introduced a framwork called SPDDL \textcolor{black}{(Secure and Privacy-preserving distributed deep learning) for secure and privacy-preserving Distributed Deep Learning (DDL)} that makes better security, efficiency, and functionality and preserves users identities from external adversaries. Likewise, a framwork that is secure and provides authentication for E-Education (online and virtual education) in cloud systems has been presented in \cite{Sfut-Conf001} to overcome cost, time processing, and security challenges. As opposed to \cite{Sfut-Conf002} that authors have indicated how to make use of several software-as-a-service (SaaS) on a platform-as-a-service (PaaS) by analyzing data security also the integration between machine learning algorithms and secure data protection algorithms has been studied in \cite{Sfut-Conf002} similar to \cite{Sfut-Conf005} that the authors produced and introduced a Phishing Detection (PD) framework that has a three-step feature and developed in machine learning and a Secure Storage Distribution (SSD) in order to provide security and privacy Cloud-IoT data. In addition, a novel firewall method called Enhanced Intrusion Detection and Classification (EIDC) has been introduced in \cite{Sfut-Conf003}. This firewall detects and ranks by a technique named most frequent decision that increases the performance of learning as opposed to \cite{Sfut-Conf004} that a model of Quality of Services (QoS) has been introduced that is new and intelligent. The critical feature in that model is deciding the files to be decrypted without decrypting the whole file list in the secure Storage. One of the most crucial concerns is balancing security and performance, which affect each other in the cloud user community. In deep learning networks on a bare metal cloud-based, a system called License Plate Recognition System (LPRS) has been constructed and introduced in \cite{Sfut-Conf006}. the purposes of this system are plate localization, character detection, and segmentation as opposed to \cite{Sfut-Conf007} that authors have tried to improve the speed and accuracy of facial recognition without losing the privacy of robots in the encrypted domain. Similar to \cite{Sfut-Conf009} that a model of cloud data repository has been introduced that has trained with mobile data dataset through \textcolor{black}{ Training dataset Filtration Key Nearest Neighbor (TsF-KNN) }and public-key cryptographic algorithms. The algoritms decline the time of the process and provide privacy and integrity of data. Furthermore, the authors in \cite{Sfut-Conf008} have argued the advantages of M-Learning by utilizing Mobile cloud computing and analyzed issues and security. Also, they have provided an introduction of Mobile cloud computing structures. Likewise, the authors in \cite{Sfut-Conf010} have developed a face recognition that is secure in cloud computing in order to protect privacy by random unitary transform encryption algorithms and for dealing with noise and modeling error by making use of multi-device diversity.

\subsection{AI in Privacy-Preserving Cloud}\label{sub3}
 Although federated learning has been designed to provide data privacy, the data uploaded from federated learning has been threatened by attackers \cite{c2,c3}. In \cite{Fut-Earl-Jour026}, authors have introduced an innovative mechanism to preserve data privacy on cloud-edge learning systems similar to \cite{Fut-Non-Earl-Jour001} that the authors have introduced an encryption technique called BGV that encrypts data and makes use of cloud servers algorithms to train for the deep computation model. Furthermore, the secure-centralized-computation-privacy-preserving reinforcement learning algorithm has been presented in \cite{Fut-Non-Earl-Jour002} in order to protect the privacy of users' data by using a fully homomorphic encryption scheme in a cloud computing infrastructure. In order to provide better complexities and security, a deep neural network architecture has been introduced in \cite{Fut-Non-Earl-Jour003}, called MSCryptoNet that works based on a fully homomorphic cryptosystem in the privacy-preservation setting. Nevertheless, there is a growing concern about the privacy and security of data when it comes to machine learning. In \cite{Fut-Non-Earl-Jour004}, the authors have introduced a secure cloud-intelligent network framework that preserves the privacy of data as opposed to \cite{Fut-Non-Earl-Jour005} that authors have produced a framwork that organizes sparse coding in edge and cloud networks. In this framework for recognizing noise and error of data, they have utilized classification. Moreover, one of the most popular applications in the fog-cloud computing environment is distributed deep learning (DDL) because of its efficiency and scalability. However, there are challenges when using DLL in fog-cloud computing, such as protecting users' privacy in the training process and confirming users' identities forged by external adversaries. According to those problems, the authors in \cite{Fut-Non-Earl-Jour006} have presented a secure and privacy-preserving DDL (SPDDL) for fog-cloud computing. In addition, one of the challenges in the internet of things (IoT) is big data that have been generated. A double-projection deep computation model (DPDCM) for feature learning in big data has been introduced in \cite{Fut-Non-Earl-Jour007}, which has two layers. In addition, For training the DPDCM, the authors have designed a learning algorithm. However, our result in neural network learning depends on how much data we use in the training step. If we want to use another party's data set, we need a solution because no party wants to disclose people's private data to others. In \cite{Fut-Non-Earl-Jour008}, authors have described their practical multiparty Back-Propagation neural network learning scheme over arbitrarily partitioned data. In this model, the Cloud runs most operations on encrypted data from ciphertext uploaded. Whereas In \cite{Fut-Non-Earl-Jour009}, the authors have argued the privacy of users' data which is used to train. They introduced a model that encrypts users' data with their public key called the privacy-preserving deep learning model. Furthermore, in smart cities where urban facilities use technology such as smart health, parking, transport, We are dealing with the date which its privacy and security are vital. In \cite{Sfut-Conf012}, the authors introduced a privacy-preserving autoencoder-based deep learning classifier in the Cloud for solving this problem in smart cities. In addition, Homomorphic Secure Multi-party Computation (SMC) or Hmomorphic Encryption (HE) schemes have been used for the security of the cloud process with considering the privacy of data. The authors in \cite{Sfut-Conf013} have argued supervised and unsupervised machine learning capability through neural networks on encrypted data similar to \cite{Sfut-Conf014} that for privacy-preserving deep neural networks, authors have introduced a framwork for transformation network training in coordination. By using the framework, users can train a transformation network with a model from a cloud provider in a secure environment. Moreover, for analyzing big data, at first, organizations should remove Personally Identifiable Information (PII) because it breaks users' privacy. A Mondrian-based k-anonymity and Deep Neural Network (DNN) based framework has been presented in \cite{Sfut-Conf015} to preserve data privacy; their analysis shows that it was suitable similar to \cite{Sfut-Conf016} that in DNN training/classification a new algorithm was presented that filtered images in order to preserve cloud base privacy. in addition, the authors have presented a cloud-based federated approach that makes use of multiple clouds with different distorted data sets and demonstrated appropriate accuracy of DNN with the original dataset can be provided. In contrast, a technique that analyzes images on the client devices and is independent of local \textcolor{black}{Convolutional Neural Network (CNN)} has been presented in \cite{Sfut-Conf017}. Also, the most important effect of CNN-based applications in the mobile Cloud that works on raw image data is serious privacy concerns. Furthermore, In \cite{GGHO-Fut-Non-Earl-Jour005} the challenges in time latency, energy consumption, and privacy level have been analyzed in an Edge-Cloud Collaboration (ECC) scenario including several users. The authors presented a Markov Decision Process (MDP) for balancing cost and the privacy level and the Deep Q-Network (DQN), which decreases delay, energy cost, and improves privacy. However, In the cloud environment, data protection is a vital step. In \cite{Fut-Non-Earl-Jour010} a Machine Learning and Probabilistic Analysis based Model (MLPAM) has been introduced in order to provide security and data protection in the cloud environment. In addition, For reducing data leakage, MLPAM has a powerful sharing protocol. To that end, a growing number of organizations have utilized machine learning applications for many purposes, such as attack detection. We can make sure of its accuracy and efficiency with a privacy-aware deployment method (PDM) that has been presented in \cite{Fut-Non-Earl-Jour011}. With this method, we can provide privacy which is a challenge in cyber-physical cloud systems (CPCSs).

\subsection{Quantum-Inspired AI}\label{sub4}
Processing a vast amount of dataset that is complex requires evolving algorithms to provide a way for better training and learning. Since The trajectory planning problem is one of the challenges in a wireless uplink transmission scenario, the authors in\cite{Quant-AI-Jour001} have presented a solution that is optimizing via the quantum-inspired reinforcement learning (QiRL) method that improves the performance of an uncrewed aerial vehicle from when it starts flying to its destination. Moreover, New training algorithms have been presented for deep reinforcement learning (DRL) to equilibrium exploration and exploitation in \cite{Quant-AI-Jour002}. As opposed to the traditional mechanism, this algorithm is inspired by quantum computation. Also, in \cite{Quant-AI-Jour003} quantum-inspired multidirectional association model called QMAM and a one-shot learning and self-convergent iterative learning called IQMAM. Their analysis has indicated the ability to maintain a stable memory capacity and reliable recall in the model. In contrast, in \cite{Quant-AI-Jour004}, autonomous mobile robots have been controlled navigationally by A new Quantum-inspired Reinforcement Learning (QiRL) method. According to Markovian tests, the QiRL method is better than traditional reinforcement learning in learning and initial states. Likewise, In\cite{Quant-AI-Jour005}, authors have been presented Quantum-inspired Fuzzy Based Neural Network (Q-FNN) model that is new in learning for challenging two-class classification. This model is much better for accuracy, sensitivity, and specificity.
\subsection{Bio-inspired AI}\label{sub5}

In \cite{Fut-Bio-Jour001}, bio-inspired representation learning is said to combine low-level contrast and high-level semantic features in order to generate a visual attention map. As outlined in \cite{Fut-Bio-Jour002}, a neural network can be used to train Bio-inspired features. Some such examples can be seen in \cite{Fut-Bio-Jour003} where bio-inspired technologies are used to simulate animal movements in aerial vehicles, and in \cite{Fut-Bio-Jour005} where it is used for autonomous navigation. In \cite{Fut-Bio-Jour004}, bio-inspired metaheuristic algorithms are used for the detection of spam emails and a bio-inspired algorithm in \cite{Fut-Bio-Jour006} is used for optimizing the Internet of Things. Another optimization algorithm can be found in \cite{Fut-Bio-Jour007} where it is used for UAV planning.

\begin{figure}[h]
	    \centering
	    \includegraphics[width=0.9\linewidth,keepaspectratio]{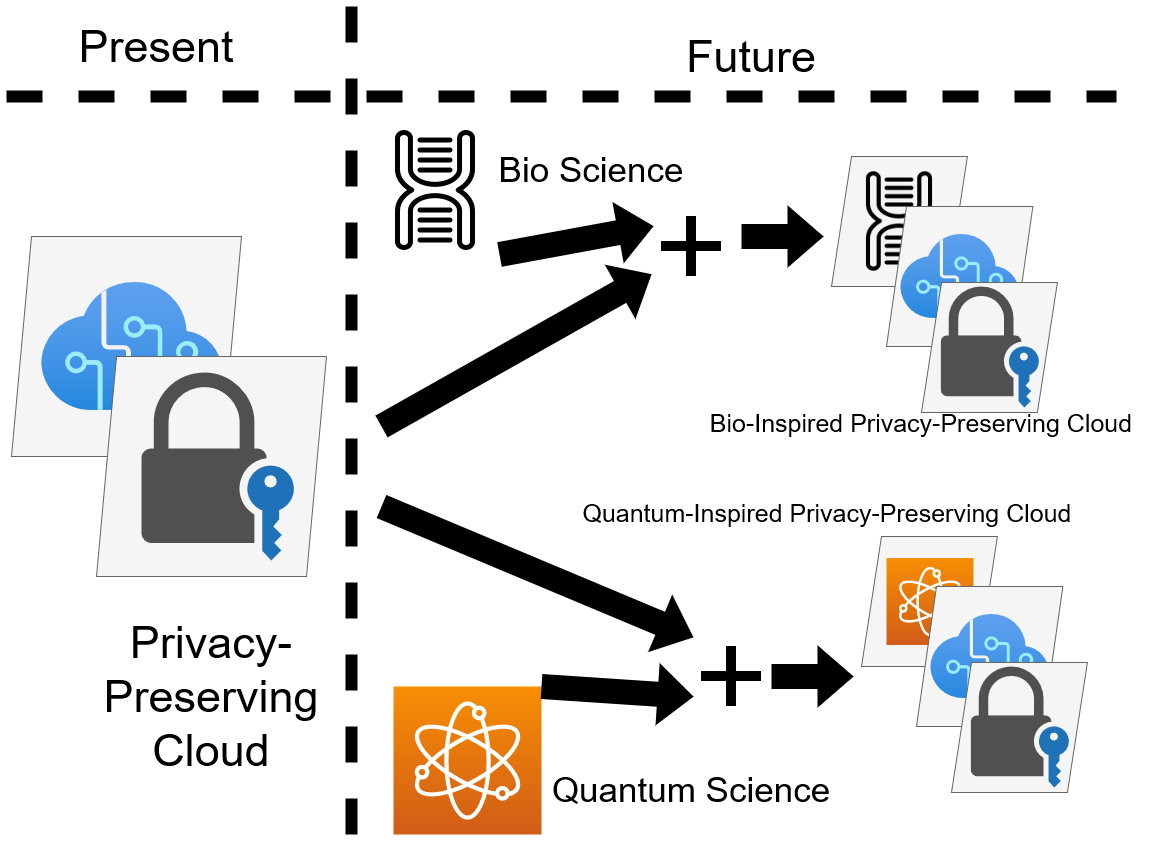}
	    \textcolor{black}{
	    \caption{Bioscience and Quantum Join AI-assisted Privacy Preserving Cloud to Create Bio-inspired and Quantum-inspired AI-assisted privacy-preserving cloud.}\label{MyFig2}}
\end{figure}

\textcolor{black}{ In Figure \ref{MyFig2}, the left side of the figure represents the current state of Privacy-Preserving Cloud technology, whereas the right side shows the future for this technology. The arrows from Privacy-Preserving Cloud point towards other technologies with which it can join to create new technologies. The DNA symbol represents Bio Science. Privacy-Preserving Cloud joins with Bio Science to create Bio-Inspired Privacy-Preserving Cloud. The Atom symbol represents Quantum Science, which joins with Privacy-Preserving Cloud to create Quantum-Inspired Privacy-Preserving Cloud.}


\section{Conclusion}
\textcolor{black}{Over the course of this survey paper, we have highlighted the Ecosystem, Applications, Architecture, and Lifecycle of Privacy-Preserving Cloud Computing. By creating this survey paper, future researchers who would like to learn more about Privacy-Preserving Cloud can now use this paper as a resource to strengthen their knowledge of the subject. Because of the benefits} that Cloud Computing has provided, it is used in many fields. However, because of the high level of usage of Cloud Computing across so many industries and technologies, there is a large amount of sensitive data that is potentially exposed. Paying attention to this subject, there is a growing concern about privacy in Cloud Computing. Having looked at this issue, Privacy-Preserving Cloud Computing technologies and methods have emerged in order to provide security and privacy. To recap the main points, current surveys from 2010 have been reviewed, and the summary of them has been provided in the tables that can be found at the end of Section Two. Those tables indicate whether the survey focuses on privacy or draws a future roadmap or the cloud computing has been reviewed in a specific application. Furthermore, Design, Verification, Implementation, and Evaluation that are categorized as the four stages of the Life Cycle of the Privacy-Preserving Cloud system have been reviewed. The layered  of a cloud computing system is also detailed in order to explain the systems that work together in order to make sure that a cloud system can stay private. Moreover, the benefits of Privacy-Preserving Cloud Computing have been observed in the ecosystem. The future road map that consists Quantum-Inspired AI and Bio-Inspired AI have been presented in this paper. \textcolor{black}{We would encourage researchers wishing to continue our work to study these areas further as the technology develops in the future.} Using these new fields, the ways in which cloud computing can be used as a tool can increase in even greater numbers.

\bibliographystyle{IEEEtran}
\bibliography{References}

\end{document}